# An FPGA-Based Hardware Accelerator for Energy-Efficient Bitmap Index Creation

**XUAN-THUAN NGUYEN**[ID]**[1,2], (Member, IEEE), TRONG-THUC HOANG[2], (Student Member, IEEE), HONG-THU NGUYEN[2], (Student Member, IEEE), KATSUMI INOUE[2,3], AND CONG-KHA PHAM[2], (Member, IEEE)**

[1] Department of Electrical and Computer Engineering, University of Toronto, Toronto, ON M5S 1A1, Canada
[2] Department of Communication Engineering and Informatics, University of Electro-Communications, Tokyo 182-8585, Japan
[3] Advanced Original Technologies Company, Chiba 277-0827, Japan

Corresponding author: Xuan-Thuan Nguyen (xuanthuan.nguyen@utoronto.ca)

**ABSTRACT** Bitmap index is recognized as a promising candidate for online analytics processing systems, because it effectively supports not only parallel processing but also complex and multi-dimensional queries. However, bitmap index creation is a time-consuming task. In this paper, by taking full advantage of massive parallel computing of field-programmable gate array (FPGA), two hardware accelerators of bitmap index creation, namely BIC64K8 and BIC32K16, are originally proposed. Each of the accelerator contains two primary components, namely an enhanced content-addressable memory and a query logic array module, which allow BIC64K8 and BIC32K16 to index 65 536 8-bit words and 32 768 16-bit words in parallel, at every clock cycle. The experimental results on an Intel Arria V 5ASTFD5 FPGA prove that at 100 MHz, BIC64K8 and BIC32K16 achieve the approximate indexing throughput of 1.43 GB/s and 1.46 GB/s, respectively. The throughputs are also proven to be stable, regardless the size of the data sets. More significantly, BIC32K16 only consumes as low as 6.76% and 3.28% of energy compared to the central-processing-unit- and graphic-processing-unit-based designs, respectively.

## I. INTRODUCTION

There has recently been massive growth in the amount of global data that is generated from web services, social media networks, and science experiments, as well as the "tsunami" of Internet-of-Things devices. According to a Cisco forecast, total data center traffic is projected to hit 15.3 zettabytes, or $15.3 \times 10^{12}$ GB, by the end of 2020 [1]. Gaining insight into the enormous ocean of data is very important because valuable data are the driving force for business decisions and processes, as well as scientists' exploration and discovery. However, analyzing such a massive amount of data is undoubtedly a crucial and time-consuming task.

As a result of this complexity, the process of data analytics is composed of many stages. To begin the process, data from different sources are collected, cleaned, and transformed into a defined format, so they then can be stored and loaded into an analytics system, such as a data warehouse. The data index and data query stage are performed, whenever users want to query information. The data index stage is usually completed in advance to reduce the execution time of the data query stage. Finally, the query results are appropriately visualized to aid business executives or end users in their decision making.

Efficient approaches to index and query are indispensable to constructing a high-performance data-analytics system. Specifically, an efficient index method aims to look any data up in the database as fast as possible, while an efficient query method should use the index results to answer all queries as fast as possible. The concept of indexes in a database management system (DBMS) is similar to book catalogs in a library or even an index in a book. Depending on the applications, several indexing frameworks have been proposed, such as B-tree index and hash index. Among them, bitmap index (BI) plays a prominent role in the solutions of highly complex and multi-dimensional queries, which usually occur in the online analytics processing systems [2].

BI was first defined by Wong *et al.* [11] and was later popularized by O'Neil and Quass [12]. A BI is a bit-level matrix, where the number of rows and columns are the length and cardinality of the data sets, respectively. With a BI, answering multi-dimensional queries becomes a series of bitwise operators, e.g., AND, OR, XOR, and NOT, on bit columns.

As a result, a BI has proven to be very useful for solving complex queries in large enterprise and scientific databases. More significantly, because of the usage of low-hardware logical operators, a BI appears to be suitable for advanced parallel-processing platforms, such as multi-core central processing units (CPUs) [13]–[16] and graphics processing units (GPUs) [17].

Modern field-programmable gate arrays (FPGAs) have become increasingly important in data analytics because of their advantages in compute-intensive tasks and energy consumption. Since 2014, Microsoft has integrated the FPGAs into 1,632 servers for accelerating the Bing search engine [21]. The experiments showed that this system achieved a 95% improvement in throughput at each ranking server with an equal latency distribution, as compared to pure software implementation. Another example is a large scale deep neural network system [22], where a mid-range FPGA could process around 3.5 and 2.5 times faster than CPU and GPU, while it consumed less than 20 W, which is about four and 22 times lower than that of the 80-W CPU and 225-W GPU, respectively.

Based on that observation, this research aims to exploit the advantages of FPGA to accelerate the creation of BI. This is because such a task is widely considered as the most expensive task in query processing. In fact, according to J. Chou *et al.* [13], more than two hours were taken to build the necessary BIs from a 50-TB data set, whereas only 12 seconds were required to answer a query using those generated BIs. The contribution of our work includes:

- The relation between a random-access-memory-based content-addressable memory (RAM-based CAM or R-CAM) and BI. We then propose a methodology to index data by borrowing the concept of R-CAM.
- The enhanced architecture of R-CAM using the bit-sliced technique to benefit greatly from the large bus width of modern systems. Concretely, if the system bus width is 256 bits, the time to reset CAM and load data to CAM go down 32 fold.
- The parallel architecture of a query logic array module containing a set of logic gates and multiplexer. This module allows the range index of a large number of data, i.e. 65,536 8-bit words or 32,768 16-bit words, to be done simultaneously.
- The full design of BI creator (BIC) that can directly access the DDR3 memory to obtain the data and keys. After completing the indexing tasks, BIC returns all of the indexes to DDR3 memory directly and begins the next indexing tasks.

Two versions of BIC, namely BIC64K8 and BIC32K16, were implemented in an Intel Arria V FPGA development board (formerly Altera) [23], which contains two separate 1-GB DDR3 memories and one Arria 5ASTFD5K3F40I3 FPGA. Roughly speaking, BIC64K8 and BIC32K16 could index as many as 65,536 8-bit words (cardinality = $2^8 = 256$) and 32,768 16-bit words (cardinality = $2^{16} = 65,536$) in parallel, respectively. The R-CAMs are easily scalable by adding or removing correspondent logic and memory resources. Various test data and keys were generated to verify two BIC's performance. The experimental results indicated that at 100 MHz, BIC64K8 and BIC32K16 produce the indexing throughput up to 1.43 GB/s and 1.46 GB/s, respectively. The throughputs are also proven to be stable, regardless of the size of data sets. More significantly, the energy consumption of BIC32K16 only holds about 6.76% of a CPU-based design [16] and 3.28% of a GPU-based design [17].

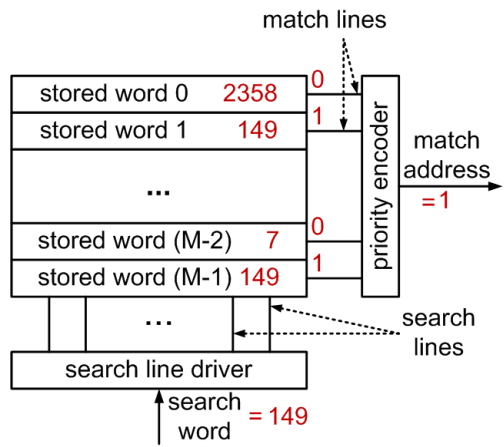

**FIGURE 1.** The simplified architecture of CAM.

The remainder of this paper is organized as follows. Section II presents the background of R-CAM and BI. Section III describes in detail the hardware architecture of BIC. Section IV gives the experimental results of BIC in an Intel Arria V FPGA development kit. Section V discusses several limitations of the proposed BIC. Finally, Section VI concludes this study.

## II. BACKGROUND
### A. RAM-BASED CAM
#### 1) OVERVIEW
CAM is a particular type of computer memory that is applied in various search-intensive applications, such as multimedia processing, data analytics, and data mining [3]. In contrast to RAM, each input and output of CAM is the content of data and address of matching data, respectively. Depending on different applications, either binary CAM (BCAM) or ternary CAM (TCAM) is used. The former only supports storage and searching binary bits (zero or one), while the latter allows a third matching state, called "X" or "don't care", in its storage.

Fig. 1 shows a simplified block diagram of a CAM. The input to the system is the search word that is broadcast onto the search lines to all CAM cells. Each stored word has a match line that indicates whether the search word and stored word are different (mismatched) or identical (matched). The match lines are fed to a priority encoder to produce the matching location, or CAM address. In short, the overall function of a CAM is to take a search word and return the matching address. Taking an example in Fig. 1, if the content of CAM is {2358, 149, . . . , 7, 149} and the search data is 149, the match lines, where the stored words are 149, turn into one (01 . . . 01). The priority encoder then selects the highest priority matching bit, i.e. words in lower address locations receiving higher priority, and encodes it into the binary value (match address = 1).

Although modern FPGAs provide a large number of embedded RAM blocks, dedicated registers, and lookup tables, they exclude dedicated CAM blocks, presumably because of their disadvantages of area and power. Instead, FPGA vendors propose the methodology to construct a scalable binary CAM from the available embedded RAM blocks of FPGA by performing a special mapping technique to input data and address [4], [5]. This specific CAM is called as a RAM-based CAM or R-CAM. Due to the mapping technique, the R-CAM requires 32-bit RAM for every 1-bit CAM and offers the search rate of one key per clock cycle, or the time complexity of O(*1*).

### 2) RELATED WORKS

Ullah *et al.* [6] presented a hybrid partitioned RAM-based TCAM (HP-TCAM) that first dissected the conventional TCAM table into $m \times n$ number of TCAM sub-tables. All of the sub-tables were then processed to be stored in their corresponding RAM units. The experiment in a Xilinx Virtex-5 FPGA proved that the search latency of $512 \times 36$ HP-TCAM was five clock cycles in the case of $n = 3$ and $m = 4$. The authors also conducted an enhanced architecture of HP-TCAM, namely E-TCAM [7], that reduced 33.33% of block RAMs as well as improved 63.03% of latency. An ASIC implementation of HP-TCAM, namely Z-TCAM [8], was performed in 180-nm technology and achieved a search latency of three clock cycles.

Ahmed *et al.* [9] proposed another architecture called resource-efficient RAM-based TCAM (REST), which was partially based on the principle of HP-TCAM. This design made use of the so-called virtual blocks in the RAM unit to enhance the emulated TCAM bits at the cost of reduced throughput. As the number of virtual blocks $m$ increased, the emulated TCAM bits also increased in REST by exhaustively consuming all bits available in the RAM. In fact, a $72 \times 28$-bit REST implemented in a Xilinx Kintex-7 FPGA only required 3.5% and 25.3% of memory resources, compared with HP-TCAM and Z-TCAM, respectively. Additionally, the search latency cost five clock cycles in the case of $m = 4$.

Abdelhadi and Lemieux [10] also introduced an efficient and modular technique for constructing BCAM using RAM block in an FPGA. This method names indirectly indexed two-dimensional (II2D) BCAM. In contrast to the traditional approach, the proposed one grouped addresses into sets and maintained a single pattern match indicator for each set. The match indicator for every single address was then efficiently regenerated by storing indirect indices for address match indicators. The experimental results confirmed that the II2D-BCAM could reach 153-bit pattern width for $m = 16K$ entries in an Intel Stratix V FPGA, which was up to four times wider than traditional BCAM. Furthermore, the search latency was four clock cycles in the case of $m = 16K$.

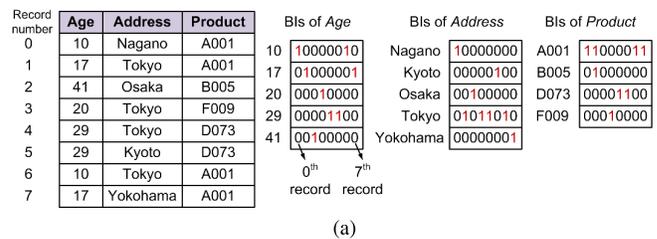

(a)

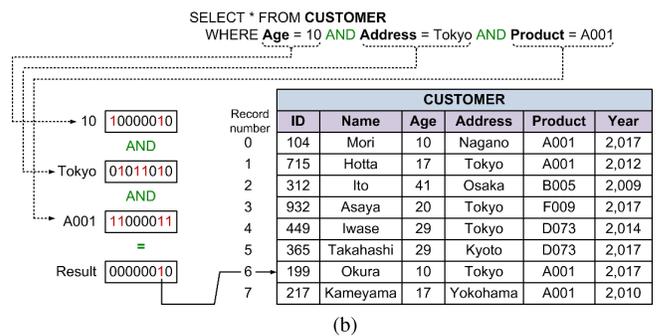

(b)

**FIGURE 2.** An example of BI-based multi-dimensional query processing. (a) The BIs of *Age*, *Address*, and *Product*. (b) The BI-based query processing.

### B. BITMAP INDEX
#### 1) OVERVIEW

BI was originally described by Wong *et al.* [11] in 1985 and later popularized by O'Neil and Quass [12] in 1997. It has proven very to be very useful for solving multi-dimensional queries in both enterprise databases and scientific databases. Fig. 2 gives an example of a relational DBMS that is comprised of a 8-record CUSTOMER *relation*. Each record, or row, stores six *attributes*, namely ID, Name, Age, Address, Product, and Year. The example query is given as "*find all 10-year-old customers who are living in Tokyo and bought product A001*". BI can be utilized to solve this three-dimensional query effectively.

Fig. 2(a) illustrates all BIs of three attributes *Age*, *Address*, and *Product*. To begin indexing, all records in CUSTOMER relation must be numbered sequentially. For each attribute, several BIs are then created. A BI is simply an array of bits. For instance, because both the $0^{th}$ record and $6^{th}$ record contain (*Age* = {10}), the $0^{th}$ bit and $6^{th}$ bit of the BI of (*Age* = {10}) turn into ones accordingly. Likewise, the $0^{th}$ bit of the BI of (*Address* = {Nagano}) becomes one, because $0^{th}$ record contains (*Address* = {Nagano}). In this way, up to 14 BIs can be created from those three attributes.

Fig. 2(b) indicates how powerfully BI can cope with the multi-dimensional query given above. To answer this query, we fetch the BIs for (*Age* = {10}), (*Address* = {Tokyo}), and (*Product* = {A001}), and then perform the bitwise logical

AND of three BIs. The intersection of three BIs gives the result of 00000010. Based on the value of the bitmap result, the $6^{th}$ record contains the result for the query because of the value of the $6^{th}$ bit. In other words, we compute a new bitmap where $i^{th}$ bit has a value one, if the $i^{th}$ bit of the three BIs are both one, and has a value zero otherwise. However, this work only deals with the BI creation because it is considered to be the most time-consuming task.

### 2) RELATED WORKS

Blanas et al. [14] introduced a new scientific DBMS on a supercomputer using three innovations. First, a relational query interface was specifically designed so that DBMS could directly access data stored in the HDF5 format without first converting and loading the data. Second, the query processing could be accelerated by simply requesting more CPU cores when the batch job was submitted for execution. Finally, BI was employed to significantly reduce the query response time. All experiments were conducted in 1,202 computer nodes, each of which contained two Intel quad-core 2.67-GHz CPU. The results proved that the proposed DBMS outperformed PostgreSQL and was more than ten times faster than Apache Hive, whenever the number of CPU cores exceeds 512 cores.

Chong et al. [15] also accelerated the BI construction of massive astronomical data sets by using a distributed environment. The distribution of data, hardware configuration, and network transmission rates were properly configured to minimize the indexing time. The experiments were conducted in a master node and three slave nodes. The master node contains an Intel quad-core 2.3-GHz CPU, while each slave node has an Intel 2-core CPU. The results showed that the indexing throughput reached 62.5 MB/s, which is about 27.9% as high as that of FastDB, a database system developed by Advanced Computing and Medical Information Services Engineering Laboratory.

Hsuan-Te et al. [16] took advantage of the growing memory capacity on a supercomputer to extend their previous work [13] to an in-memory query system for scientific data analysis. The system combined four techniques, namely bitmap indexing, spatial data layout reorganization, distributed shared memory, and location-aware parallel execution, for maximizing the processing. Specifically, first two techniques accelerated queries with range and spatial constraint across multiple data sets, while the rest provided the capability of caching and transforming data sets for data analysis. All experiments were conducted in 5,576 compute nodes, each of which contained two Intel 12-core 2.4-GHz CPUs. The experimental results showed that the proposed system achieved a throughput of 28 GB/s and 510 GB/s, when using 1,250 cores and 20,000 cores, respectively.

Fusco et al. [17] implemented two compressed BI creation systems on an NVIDIA Geforce GTX 670 GPU, which targets the multi-10-Gbps network traffic recorders. In contrast to CPUs, which rely on large caches to hide memory latencies, GPUs are optimized for throughput by distributing the computation into thousands of hardware threads executed in parallel. The authors exploited such power by first copying all numerical data from the host memory to the GPU. The GPU computed and returned a serialized BI to the host memory together with the information required to access individual index columns. Regarding the indexing throughput, the GPU-based design could index 20 million random 16-bit numbers within 122.6 $\mu s$, or the throughput reached 163 million words/second.

Our previous work [18] combined a R-CAM with a bit-level transpose matrix module to index a $N$-record document with $M$ keys. In the beginning, record zeroth was loaded into R-CAM. Subsequently, $M$ keys were dispatched to R-CAM. The outcome of this phase was a $M$-bit BI array, where bit $j$ indicated the existence of key $j$ in the current record, and vice versa. The process repeated until $N$ records were completely indexed. Afterwards, the $N{\times}M$-bit BI matrix was transposed into a $M{\times}N$-bit BI matrix, which could be further processed by a DBMS. The experiments in an Intel Cyclone V SX FPGA showed that if $N = 256$ records and $M = 16$ keys, the indexing throughput would have reached 5.7 million records per second at 150 MHz. An application-specific integrated circuit (ASIC) implementation was also verified successfully in a 65-nm Silicon-On-Thin-Buried-oxide (SOTB) CMOS process [19].

Our most recent work [20] proposed a regular expression matching based on BI, instead of finite-state automation, for text analytics. In the beginning, all text data were loaded into R-CAM. A $N$-character query is then sent into R-CAM. After $N$ clock cycles, the indexes of all matches can be found in the R-CAM output. The experimental results in an Intel Arria V ST FPGA proved that our design could process a 64-character query inside a 64-KB text data within 43.76 $\mu s$ at 100 MHz, or the throughput reached 11.98 Gbps.

### C. MOTIVATION
#### 1) THE NEED FOR AN ENHANCED R-CAM

Most of the previous studies on R-CAM mainly focused on the improvement of mapping techniques, thereby helping to reduce the memory utilization [6]–[10]. In spite of the remarkable achievement in memory reduction, those designs contained two main drawbacks that fatally affect the high-throughput indexing system. First, the search latency heavily counted on the partitioning settings and the R-CAM size, and took more than one clock cycle to complete. Second, an effective mechanism for loading data into R-CAM could not be resolved. In fact, those designs could only receive each word per clock cycle, so they could not fully exploit the high-bandwidth bus in the modern systems. For example, if the bus width contains 256 bits, an R-CAM requires at least 32 clock cycles to load all of the 32 8-bit words. Such update latency unquestionably degrades the entire system performance. For those reasons, an enhanced R-CAM, which can load data at system-like bandwidth and attain the search rate of one key per clock cycle, is essential for the high-throughput indexing system.

### 2) THE NEED FOR AN FPGA-BASED BI CREATION

Since BI creation is widely considered as the most expensive task in query processing, reducing the indexing time has attracted researcher's attention. Although CPU-based systems [13]–[16] and GPU-based systems [17] can improve the indexing time due to their parallel processing capabilities, the power consumption increases proportionally to the resource usage, e.g., the number of used CPU and GPU cores. Hence, there is an upward trend towards new platforms, such as FPGAs, with the target of energy efficiency. Energy efficiency is defined as the ability to deliver the same processing throughput, but consume less power. For instance, the conventional CPUs consume a large amount of energy and cannot be optimized to suit the target applications. On the other hand, GPUs are programmable, but use an even higher amount of energy. FPGAs, however, offer a middle ground among the platforms with high energy efficiency without sacrificing the throughput of the application.

Even though the concept of using R-CAM for indexing has been proposed in our previous work [18]–[20], the aims and objectives were vastly different. In fact, those works focused on full-text search applications. In [18] and [19], R-CAM only returned one or zero, which determined whether a given text record contain a certain key list. It is different from this work, where a BI vector was returned for each key value. Each bit of this vector indicated the presence of matching. In [20], R-CAM was employed to speed up the regular express matching process. Despite being originally introduced to improve the R-CAM loading latency, the detailed description of bit-sliced technique was not provided. The impact of this technique on indexing throughput was also not mentioned. This work, therefore, aims to address those questions. Additionally, an effective BIC architecture, which provides both point indexes, e.g. BI of ($Age = \{10\}$), and range indexes, e.g. BI of ($Age \neq \{10, 17, 29\}$) or BI of ($Age \leq 10$), are proposed.

The uncompressed BI is concentrated because the outcome of BIC will be processed directly by a BI-based query processor [27]. In other words, the processor receives the raw BIs generated by BIC to answer the given multi-dimensional queries. In case of the example in Fig. 2(b), three BIs of ($Age = \{10\}$), ($Address = \{Tokyo\}$), and ($Product = \{A001\}$) are first copied to this processor. The processor then performs two logical operations and delivers the results at clock cycle rate. The previous works proved that the processor was fully operational at 50 MHz and achieved the processing throughput of 32-Kbit BI/operation/cycle in an Intel Arria V ST FPGA.

## III. PROPOSED SYSTEM ARCHITECTURE
### A. OVERVIEW

Fig. 3 illustrates the block diagram of a BIC used to index three given attributes (columns), namely *Age*, *Address*, and

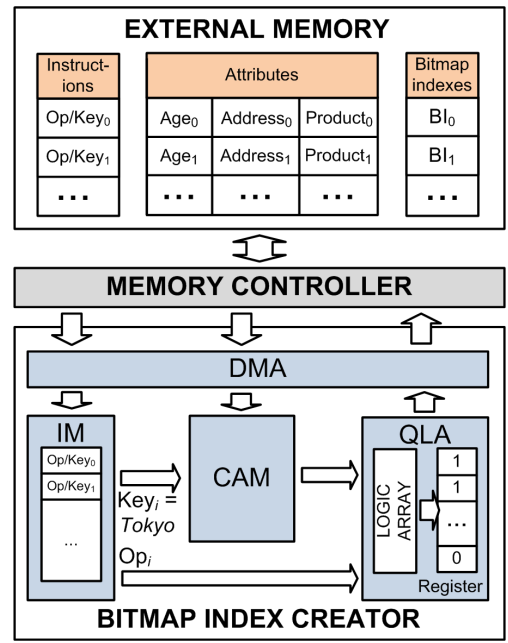

**FIGURE 3.** The block diagram of BIC.

*Product*, of CUSTOMER relation. All of the instructions and keys are extracted from the queries and initially stored in an external memory together with all attributes. BIC is composed of four modules operating in parallel, namely a direct memory access (DMA), a RAM-based CAM, an instruction memory (IM), and a query logic array (QLA).

For instance, to index attribute *Age*, all operation/key values are first transferred to the IM. Subsequently, R-CAM continuously receives the values of *Age* until R-CAM becomes full. R-CAM then starts using the operation/key (Op/Key) values from IM to produce the correspondent BIs. Each separate BI is dispatched to QLA in turn, where an array of logic gates and an internal register are employed to calculate the range BIs. Finally, the register values or range indexes are orderly stored in the external memory. This process repeats until all values of the current attribute are indexed. The next two attributes also follow the same process.

### B. DIRECT ACCESS MEMORY (DMA)

The Arria V SX FPGA provides the high-throughput memory controllers so that users can exploit the considerable space of external memory. Because the currently used DDR3 memory only supports the theoretical bandwidth of 25.6 Gbps, the bus data width and the operating frequency of the memory controller are configured 256 bits and 100 MHz, respectively. The three-channel DMA module is adequately designed so that BIC can directly access data in DDR3 memory up to the theoretical bandwidth.

### C. RAM-BASED CAM (R-CAM)
#### 1) RELATIONSHIP BETWEEN BITMAP INDEX AND R-CAM

As mentioned earlier, modern FPGAs exclude dedicated CAMs because of their disadvantages of area and power.

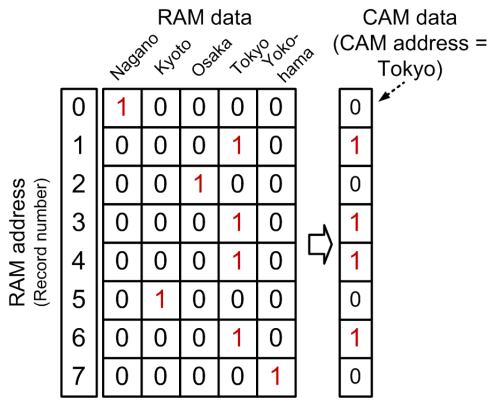

**FIGURE 4.** The block diagram of R-CAM of *Address*.

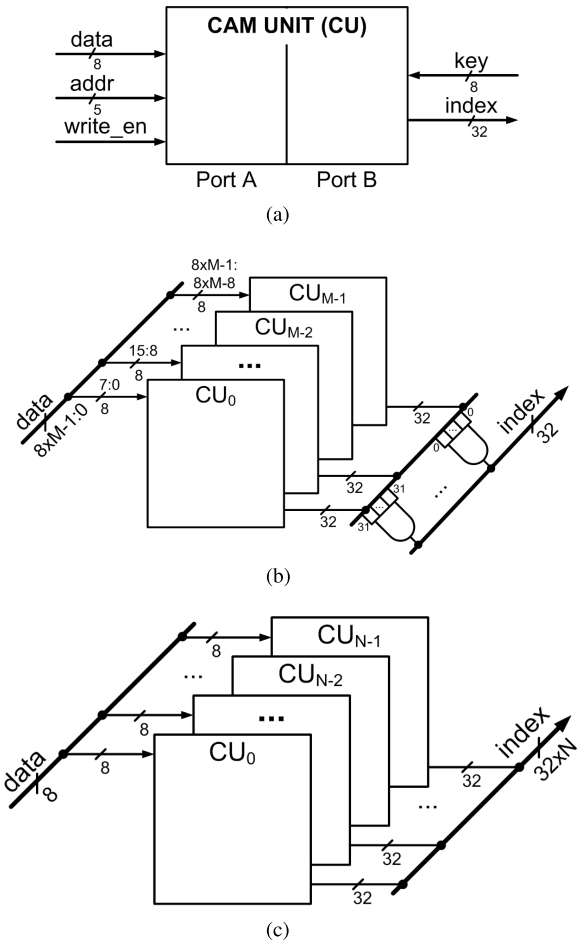

**FIGURE 5.** (a) A 32×8-bit R-CAM. (b) A 32×(8×M)-bit R-CAM. (c) A (32×N)×8-bit R-CAM. (d) The block diagram of cascade R-CAMs.

Instead, CAMs are built from the embedded RAM blocks of FPGA by applying a special mapping technique. Fig 4 depicts a R-CAM of attribute *Address* that employs the mapping technique. Each row in the RAM represents one possible mapping of the input data bits to the R-CAM contents. Concretely, the value of each cell is set to one, if the data is stored at that address, and vice versa. Intuitively, R-CAM mentioned in this example is a transpose of the BIs of *Address*. In fact, the RAM value at address $j$ is equivalent to a BI vector $j$, or column $j$ of BI matrix depicted in Fig. 2(a). For this reason, the concept of R-CAM is borrowed to construct a BIC.

### 2) SIMPLIFIED ARCHITECTURE OF R-CAM
This section briefly summarizes the architecture of traditional R-CAM described in [4] and [5], from which the enhanced architecture is proposed. An R-CAM is built from the dual-port memory (DPM). Specifically, the Intel Arria V FPGA exploits a DPM, where port A is an 8,192×1-bit memory and port B is 256×32-bit memory, to construct a 32×8-bit R-CAM, also named as R-CAM unit (CU). One R-CAM bit, therefore, costs 32 RAM bits. These settings are selected because 32×8-bit R-CAM is the most efficient R-CAM primitive that can be built from a basic memory block unit M10K of the Arria V FPGA.

Fig. 5(a) describes the architecture of a CU, whose input data *data* and addresses *addr* enter at port A, while search data (*key*) and matching address (*index*) leave at port B. R-CAM is scalable by connecting the CUs horizontally or vertically. Fig. 5(b) shows a 32×(8×M)-bit R-CAM, where the input data are divided into M segments and each of them enters each correspondent CU simultaneously. The output can be seen as 32 M-bit groups and each connects to an M-bit AND gate. Another illustration of a (32×N)× 8-bit R-CAM is shown in Fig. 5(c), where the input data are put into each CU in turn from CU$_0$ to CU$_{N-1}$. Finally, a (32×N)×(8×M)-bit R-CAM can be obtained by combining the two architectures above.

With the abundant logic elements and M10K memory blocks, an Arria V FPGA can afford as large as 65,536×8-bit R-CAM (CAM64K8) or 32,768×16-bit

R-CAM (CAM32K16). As a result of an 8-bit word and 16-bit word, CAM64K8 and CAM32K16 can support cardinality of 256 and 65,536, respectively. In addition, the update process costs two clock cycles per word, whereas the search process costs one clock cycle per BI. Take CAM64K8 as an example, because input data are fed sequentially and as many as 2×65,536 clock cycles are required to fill up CAM64K8. In general, the update time reaches O($2N$), where $N$ is the number of R-CAM words. The larger the R-CAM size is, the longer the update time becomes. Minimizing update time, therefore, is crucial to maximize the system performance.

### 3) ENHANCED ARCHITECTURE OF R-CAM
The data bus between DMA and R-CAM is 256 bits wide, due to the settings of the memory controller mentioned earlier. If each value of the attribute is eight bits long, DMA can transfer $\frac{256}{8} = 32$ values to CAM64K8 simultaneously in every cycle. Accordingly, CAM64K8 is capable of loading all values within $\frac{65,536}{32} = 2,048$ cycles. This is achieved by applying bit-sliced technique in CAM64K8, where all inputs and outputs of CUs are grouped in a specific order. Fig. 6 illustrates the architecture, where CAM64K8 is formed for resetting old data and RESET = 0 requests for loading new data. The CAM64K8 is expressed as a three-dimensional array CAM64K8[64][32][32], where the first, second, and third dimensions correspond to the number of CBs, the number of CUs in each CB, and the number of words in each CU, respectively. Two **for** loops in third and fourth lines emulate the selection of CBs, CUs, and CU words. The codes from the fifth line to seventh line are performed in parallel. If RESET = 0, the word $j$ of all 32 CUs in CB[$i$] are reset; otherwise, all of the CUs receive new data. Their function calls are described in 12th and 13th lines.

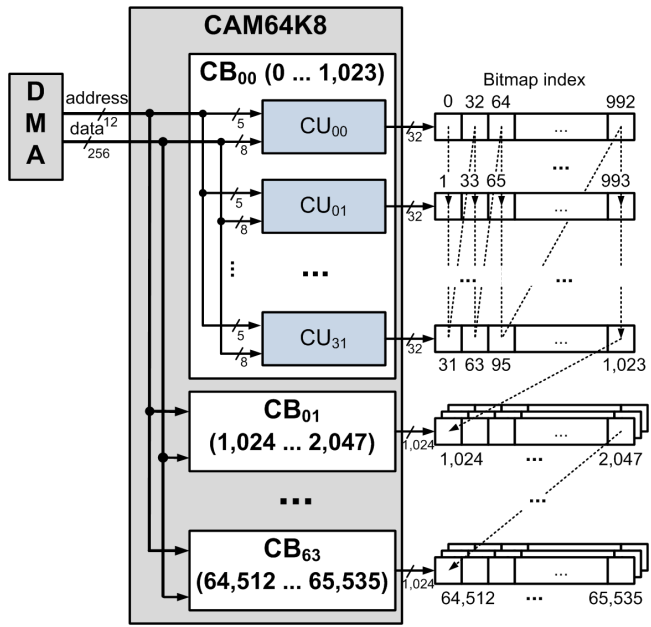

**FIGURE 6.** The enhanced architecture of a 65,536×8-bit R-CAM.

---

**Algorithm 1** Loading Data Into R-CAM

---

1: **function load_process** (DATA, RESET)
    *Input*: DATA[65536], RESET
    *Initialization*: CAM64K8[64][32][32], $i$, $j$, $d$
2: $d = 0$
3: **for** ($i = 0$, $i < 64$, $i = i + 1$) **do**
4:    **for** ($j = 0$, $j < 32$, $j = j + 1$) **do**
5:       CAM64K8[$i$][0][$j$] = RESET ? 0 : DATA[$d + 0$]
6:       CAM64K8[$i$][1][$j$] = RESET ? 0 : DATA[$d + 1$]
         ...
7:       CAM64K8[$i$][31][$j$] = RESET ? 0 : DATA[$d + 31$]
8:       $d = d + 32$
9:    **end for**
10: **end for**
11: **end function**
     *Functions calls*:
12: **load_process**(old DATA, 1)
13: **load_process**(new DATA, 0)

---

by 64 CU blocks (CBs) and each block consists of 32 CUs. The number of CUs in each CB is the quotient of data bus width and value size. The outputs are also arranged in a specific order. The first 32-bit segment of $CB_{00}$ is formed by the zeroth bit of BI of $\{CU_{00}, \ldots, CU_{31}\}$. Likewise, the second 32-bit segment of $CB_{00}$ matches the first bits of BI of $\{CU_{00}, \ldots, CU_{31}\}$ and so on. The first 1,024-bit BI is obtained by those 32 segments. As a result of this strategy, CAM64K8 can receive many characters simultaneously, while the order of its 65,536-bit BI is kept unchanged.

The loading process is summarized as follows. To begin, the first 32 values are put into the first 32 consecutive CUs of $CB_{00}$ concurrently. As soon as the $CB_{00}$ is full, the incoming values are sent to the next group of $CB_{01}$. This process continues until all data are correctly stored. Accordingly, the loading time is reduced up to 32 times as compared to the traditional structure. Furthermore, because CAM64K8 has to be cleared before receiving the new value, the update process takes two clock cycles for loading 32 values to CAM64K8, or 4,096 clock cycles for filling the entire CAM64K8. In general, by employing bit-sliced technique, the required loading time becomes O($\frac{2N}{f}$), where $N$ and $f$ are the number of R-CAM words and the ratio of system bus width and R-CAM width, respectively. This implies that the loading time will sharply decrease as long as the bus width expands. Another point is that the number of CUs in each group varies with the width of the data bus and the value size. For instance, in the case of CAM32K16, each CB only contains 16 16-bit CUs.

The algorithm of CAM64K8's loading process is stated in Algorithm 1. The input includes an 65,536-byte DATA array and a RESET signal. In fact, RESET = 1 requests

## D. INSTRUCTION MEMORY (IM)

IM stores the operation/key values extracted from the user's query. It is built from embedded RAM blocks and can contain as many as 4,096 32-bit operations. Larger IM is also easily constructed by adding more RAM blocks. Each operation is composed of two parts, as seen in Fig. 7(a). The first part is a 16-bit key value that supports the highest cardinality of 65,536. This key value can also easily be expanded to 24 bits by using the reserved bits. The second part is a 3-bit operation value that supports three operations, namely *OR*, *NO*, and *EQ*. Except *OR* and *NO*, which are the logical operations, *EQ* is only asserted whenever we transfer BI to the external memory. Due to the data width of 256 bits, up to eight 32-bit operations are loaded into IM at every clock cycle.

Fig. 7(b) gives an example of translation from a query to the correspondent operation/key values. Due to the light workload, the translation is performed in advance, such as by a computer, and the final binary configuration will be transferred to the external memory. To begin the translation, three key values and four operations are obtained from the given query. Accordingly, five opcodes are required to answer this query. Specifically, BI of ($Age = \{10\}$) is combined with the result by the *OR* operation. After three *OR* executions, the range index is reversed by the *NO* operation. The result is finally sent out by the *EQ* operation. A total of five instructions are required for this index request. It is noted that the result is automatically cleared, as soon as BIC is powered up or the *EQ* operation is completed.

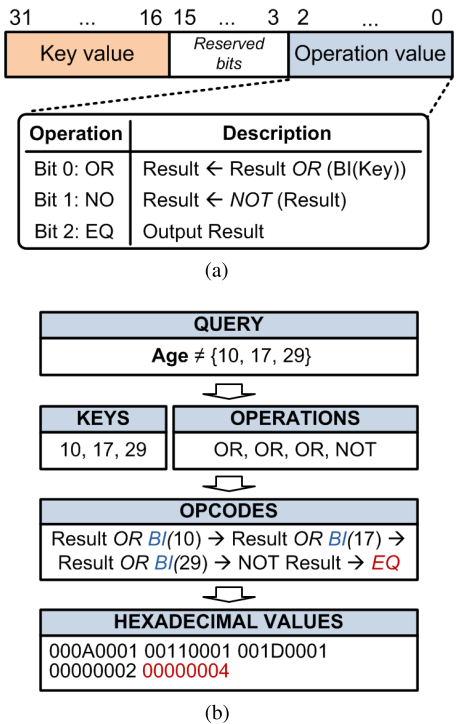

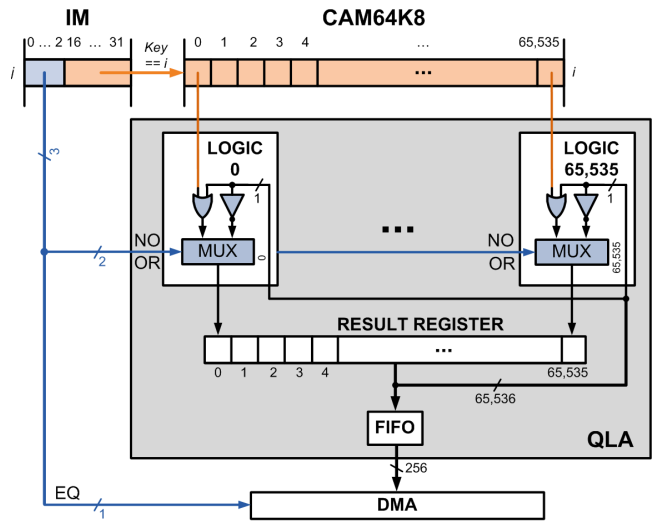

**FIGURE 8.** The block diagram of QLA.

**TABLE 1.** Notation of two proposed designs.

| Notation | Description |
|---|---|
| $N$ | The number of R-CAM words, $N = 65,536$ in BIC64K8 and $N = 32,768$ in BIC32K16 |
| $M$ | The word size, $M = 8$ bits in BIC64K8 and $M = 16$ bits in BIC32K16 |
| $N_i$ | The number of instructions (operation/key values) of IM, $N_i \leq 4,096$ |
| $B$ | The number of batches, 1 batch = 64 KB, $B \leq 8,192$ |
| $w$ | The data bus width, $w = 256$ bits |

**FIGURE 7.** The illustration of IM. (a) The structure of an IM word. (b) The example of query-to-operation/key conversion.

### E. QUERY LOGIC ARRAY (QLA)

QLA is the most compute-intensive module where the range indexes are calculated. Fig. 8 depicts the scalable architecture of QLA collaborating with CAM64K8 that contains a 65,536-bit result register and an array of the set of logic gates, including an inverter gate, an OR gate, and a multiplexer. Each bit of R-CAM output is connected with each logic set numbered from zero to 65,535. The output of each logic set also enters the result register for temporarily storage. Taking IM($i$) as an example, the 16-bit key value selects the proper BI vector in CAM64K8, while the 3-bit operation configures the multiplexers and DMA. As a result of the simplicity of the logic sets, each logical operation can be solved within one clock cycle. However, the execution time of the *EQ* operation varies with the register size. For example, in the case of CAM64K8, it takes at least 256 cycles to transfer a 65,536-bit result to the external memory. Similarly, in the case of CAM32K16, because the size of the register and logic gates are halved, only 128 cycles are needed.

The main advantage of QLA is that range indexing can be done at the rate of the clock cycle. For instance, we can obtain the BI of ($Age \neq \{10, 17, 29\}$) of 65,536 values only within four clock cycles, or 40 ns in the case of 100-MHz operating frequency. Similarly, the BI of ($Age \leq 10$) is calculated within ten clock cycles with the assumption that the smallest *Age* is one and *Age* is in integer value. Concretely, the BI of ($Age \leq 10$) is obtained by performing BI($Age = 0$) OR BI($Age = 1$) OR ... BI($Age = 10$). On the other hand, the BI of ($Age > 10$) can be calculated by putting the bitwise logical NOT after the BI of ($Age \leq 10$).

A FIFO is used to enhance the parallelism between the indexing and transfer process. If *EQ* is asserted, the range index result enters this FIFO. When the result register is fully stored in the FIFO, the next indexing process is started immediately. Unless the FIFO is empty, the DMA transfers data from FIFO to external memory. As a result, the indexing and transfer process can be operated in parallel to save the whole indexing time. Through those achievements, the proposed BIC is likely to be far better than that implemented in software.

### IV. PERFORMANCE ANALYSIS

The performance is evaluated by the hardware utilization, processing throughput, and energy efficiency of BIC when indexing a large data set under various scenarios. Table 1 summarizes the notation that is used.

**TABLE 2. The synthetic data sets.**

| | # of batches ($B$) | Size (KB) |
|---|---|---|
| DS1(8), DS1(16) | 1 | 64 |
| DS2(8), DS2(16) | 16 | 1,024 |
| DS3(8), DS3(16) | 256 | 16,384 |
| DS4(8), DS4(16) | 4,096 | 262,144 |
| DS5(8), DS5(16) | 8,192 | 524,288 |

## A. TESTBEDS

We conducted all of the experiments on an Arria V SoC development kit [23], which contains an Arria V 5ASTFD5K3F40I3 FPGA chip and two 1-GB 533-MHz DDR3 memories. The FPGA contains several hard memory controllers that allow the FPGA to access those DDR3 memories at the theoretical bandwidth of 25.6 Gbps. Hence, to fully exploit the memory bandwidth, the data bus width $w$ and the frequency of each controller are configured as 256 bits and 100 MHz, respectively. The FPGA communicates with a host computer via a PCI-Express interface, whose effective bandwidth reaches as high as 11.6 Gbps.

Each experiment follows four main steps: (1) all test operation/key values and data are transferred from a host computer to DDR3 memory via a PCI-Express interface; (2) DMA copies all operation/key values and data from this DDR3 memory to IM and R-CAM, respectively; (3) BIC indexes data using given operation/key values and then sends all generated BIs to the DDR3 memory; (4) all of the results are returned to the host computer for verification. Because the step (1) is for initialization and step (4) is for verification, the indexing process only counts (2) and (3), whose execution time is calculated by several internal counters integrated into BIC.

### 1) DATA SETS

The data sets are derived from the CUSTOMER and LINEITEM tables of the TPC-H database [24] with the scale factor of one. The attribute *c_nationkey* of CUSTOMER, which contains 25 unique values representing different country names, is used to verify BIC64K8 (cardinality of 256). The attribute *l_suppkey* of LINEITEM, which consists of 10,000 unique values symbolizing different supplier names, is used to verify BIC32K16 (cardinality of 65,536). Due to the scale factor of one, *c_nationkey* and *l_suppkey* contain as many as 150,000 and 6,001,215 words, respectively. To fix them to a 64-KB R-CAM, those data have to be divided into smaller batches. Concretely, each 8-bit batch is created by randomly selecting 65,536 words out of 150,000. Similarly, each 16-bit batch is created by randomly selecting 32,768 words out of 6,001,215 words. In our test, the number of batches varies from one to 8,192, corresponding to 64-KB to 512-MB data sets. Table 2 summarizes five data sets for BIC64K8 and BIC32K16, namely DSx(8) and DSx(16), where x ranges from one to five.

### 2) INSTRUCTION SETS

The synthetic instruction sets, numbered from IS1 to IS4, for point-index and range-index experiments are shown in Table 3. The point-index experiment returns the index of one key value only, whereas the range-index experiment returns the index of a set of keys. Suppose that $K$ denotes the given keys, in the point-index experiment such as IS1, $K$ contains one random key value from zero to 255. However, in the range-index experiment like IS3, $K$ contains a set of 1,024 distinct keys or 1,024 sequential keys, whose values range from zero to 65,536. Only IS1 and IS2 are used to evaluate BIC64K8 performance, because the 8-bit keys are expected to be in the range of zero to 255. In the case of BIC32K16, all four instruction sets are employed. However, since BIC can index data at the rate of one key/cycle, the indexing throughput is entirely independent of the key value. The experimental results below will prove this statement.

**TABLE 3. The synthetic instruction sets.**

| | Number of keys: {List of keys} | Number of operations: {List of operations} |
|---|---|---|
| IS1 | 1: $K \in \{0, ..., 255\}$ | 2: {OR, EQ} |
| IS2 | 128: $K \in \{0, ..., 255\}$ | 129: {OR, ..., OR, EQ} |
| IS3 | 1,024: $K \in \{0, ..., 65,535\}$ | 1,025: {OR, ..., OR, EQ} |
| IS4 | 4,096: $K \in \{0, ..., 65,535\}$ | 4,097: {OR, ..., OR, EQ} |

**TABLE 4. The hardware utilization of two BICs.**

| | BIC64K8 | BIC32K16 |
|---|---|---|
| R-CAM size ($N \times M$) | $65,536 \times 8$-bit | $32,768 \times 16$-bit |
| ALMs | 89,075 (51%) | 43,868 (25%) |
| – Lookup tables | 154,185 | 76,361 |
| – Registers | 82,161 | 43,101 |
| Memory (Mbits) | 16.14 (73%) | 16.14 (73%) |
| Frequency (MHz) | 118.3 | 129.5 |

## B. HARDWARE UTILIZATION

The place-and-route compilation was done by Quartus II 16.0 design software with the compiler setting of aggressive performance mode. This mode causes the compiler to target increased positive timing margin, increase the timing optimization effort applied during placement and routing, and enable timing-related physical synthesis optimizations to maximize design performance at a potential increase to logic area.

Table 4 shows the hardware consumption of BIC64K8 and BIC32K16. Adaptive logic modules (ALMs), a fundamental building block of Arria V FPGA, together with memory bits, are used to evaluate the resource utilization. Each ALM is composed of one 8-input combinational look-up table with four dedicated registers. The identical memory utilization of both designs is due to the similarity of the R-CAM size. Because one R-CAM bit costs 32 RAM bits, therefor 16-Mbit memory is required to build an entire 64-KB R-CAM. The other modules, such as IM and QLA, utilize the remaining 0.14-Mbit memory. Although BIC32K16 requires 32,768 AND gates for CAM32K16, its ALMs are still smaller than that of the BIC64K8 because of the reduction of its QLA module. The operating frequency of 118.3 MHz and 129.5 MHz were obtained by TimeQuest timing analyzer provided by Quartus 16.0. Nonetheless, the timing analysis

**TABLE 5.** The execution time of each BIC module.

| Time (clock cycle) | Description |
|---|---|
| $t_{IM} = \dfrac{N_i \times 32}{w}$ | Time to load operation/key values to IM |
| $t_{CAM} = \left(\dfrac{N \times M}{w}\right) \times 2$ | Time to reset CAM and load new batch data to CAM |
| $t_{QLA} = N_i$ | Time to process all operation/key values |
| $t_{OUT} = \dfrac{N}{w}$ | Time to output one $N$-bit BI |
| $T_{theo} = t_{IM} + (t_{CAM} + t_{QLA} + t_{OUT}) \times B$ | Theoretical indexing time |

results suggested that both designs could be integrated in a 100-MHz system without any problem.

### C. PROCESSING THROUGHPUT

#### 1) PREDICTION MODEL

To prepare for the evaluation of processing throughput, the formulas to calculate the execution time of all individual modules are first constructed, as shown in Table 5. Specifically, $t_{IM}$ is the time to load 32-bit instructions from the DDR3 memory to IM, $t_{CAM}$ is the time to reset and load all data sets from the DDR3 memory to R-CAM, $t_{QLA}$ is the time to process all instructions, $t_{OUT}$ is the time to return the BIs to the DDR3 memory, and $T_{theo}$ is the theoretical indexing time. Because we have to eliminate the old values in R-CAM before loading new data, $t_{CAM}$ is the sum of resetting and loading time. Moreover, the resetting and loading time are identical, therefore $t_{CAM}$ is twice the loading time $\frac{N \times M}{w}$. The theoretical indexing throughput (THR$_{theo}$) is then defined as the number of words that can be processed at every second.

#### 2) POINT-INDEX AND RANGE-INDEX EXPERIMENTS

Fig. 9(a) shows the practical indexing throughput (THR$_{prac}$) of BIC64K8 when the data sets vary from DS1(8) (64 KB) to DS5(8) (512 MB) and the instruction sets vary from IS1 (point index) to IS2 (range index). At DS1(8) and IS1, because the execution time is 45.7 $\mu$s, THR$_{prac}$ reaches approximately 1.43 billion words/second. Additionally, THR$_{prac}$ slightly increased by 0.2% when the data size varies from 64 KB to 512 MB. At DS1(8) and IS2, THR$_{prac}$ reaches approximately 1.39 billion words/second due to the execution time of 47.2 $\mu$s. Similarly, THR$_{prac}$ at DS5(8) is slightly higher than that of DS1(8). Moreover, THR$_{prac}$ of IS2 is around 2.9% as small as that of IS1, due to the difference in the number of instructions, i.e. IM1 and IM2 contain one and 129 instructions, respectively.

Fig. 9(b) illustrates the difference between THR$_{theo}$ and THR$_{prac}$ of BIC64K8 for different data sets and instruction sets. THR$_{theo}$ is calculated based on the formula in Table 5. Two main findings are obtained from this figure. First, when the data sets vary from DS1(8) to DS5(8), the difference is slightly reduced from 4.8% to 4.3%. Second, with the same data set, the difference at IS2 is slightly smaller than that at IS1. To find out the reason for this difference, the execution times of each stage are carefully analyzed.

Fig. 9(c) depicts the share of $t_{IM}$, $t_{CAM}$, $t_{QLA}$, and $t_{OUT}$ in the practical indexing time (T$_{prac}$) at two instruction sets and data set DS1(8). Because $t_{QLA}$ produces each BI per clock cycle, $t_{QLA}$ only depends on the number of instructions $N_i$. Taking IS2 as an example, 129 clock cycles are required to calculate the final BI. However, $t_{IM}$, $t_{CAM}$, and $t_{OUT}$ reply on the access time between FPGA and the DDR3 memory. Among them, $t_{CAM}$ contributes the highest portion.

The difference between theoretical and practical bandwidth is due to two reasons. The first reason is the latency of transition stage, which causes nearly 1% of the difference. The second reason is the memory access latency. Generally speaking, DDR3 memory is constructed by several memory banks, each one containing many columns and rows. To read or write data in DDR3, the memory controller first opens a specific row in a particular bank. The entire row of the memory array is then transferred into the corresponding row buffer. On completion, a column access command is performed to read or write data from or to row buffer. Finally, the row buffer must be written back to the memory array by a precharge command so that this bank is available for a subsequent row activation. When data sets and instruction sets increase, the memory controller optimizes the access requests, thereby slightly reducing the access time. According to the prediction model presented in Table 5, $t_{QLA}$ is also equal to $N_i$. In other words, $t_{QLA}$ is independent from the access of DDR3 memory. Hence, the increase in $N_i$ causes the decrease in the difference of IS2, as compared to IS1. Moreover, when $B$ increase, the difference is reduced because BIC accesses memory more frequently, therefore the waiting stage of DDR3 memory is reduced correspondingly. That small difference allows us to predict the access time by using the proposed prediction model.

Fig. 9(d) depicts THR$_{prac}$ of BIC32K16 at five data sets, from DS1(16) to DS5(16), and four instructions sets, from IS1 to IS4. It can be seen that the throughput is halved because the word size is doubled, from eight bits to 16 bits. It can be seen that THR$_{prac}$ are approximately 0.73, 0.71, 0.58, and 0.36 billion words/second at IS1, IS2, IS3, and IS4, respectively. As similar to BIC64K8, THR$_{prac}$ slightly increases when the sizes of the data sets increase. This can be seen clearly in Fig. 9(e), where the difference between THR$_{theo}$ and THR$_{prac}$ is slightly reduced. The difference of IS4 is the smallest because of the largest value of $N_i$. In fact, at DS5(16) and IS4, the difference comes closer to 2%. Fig. 9(f) illustrates the distribution of $t_{IM}$, $t_{CAM}$, $t_{QLA}$, and $t_{OUT}$ at DS1(16) and four instruction sets. At IS4, $t_{QLA}$ is almost equal to $t_{CAM}$, due to a large number of instructions.

In summary, the experiments above proved that at a 100-MHz operating frequency BIC64K8 and BIC32K16 achieved the maximum throughput of 1.43 billion and 0.73 billion words/second, or 1.43 GB/s and 1.46 GB/s, respectively. Moreover, because the indexing throughput was almost stable regardless the size of data sets, they can be estimated using the prediction model.

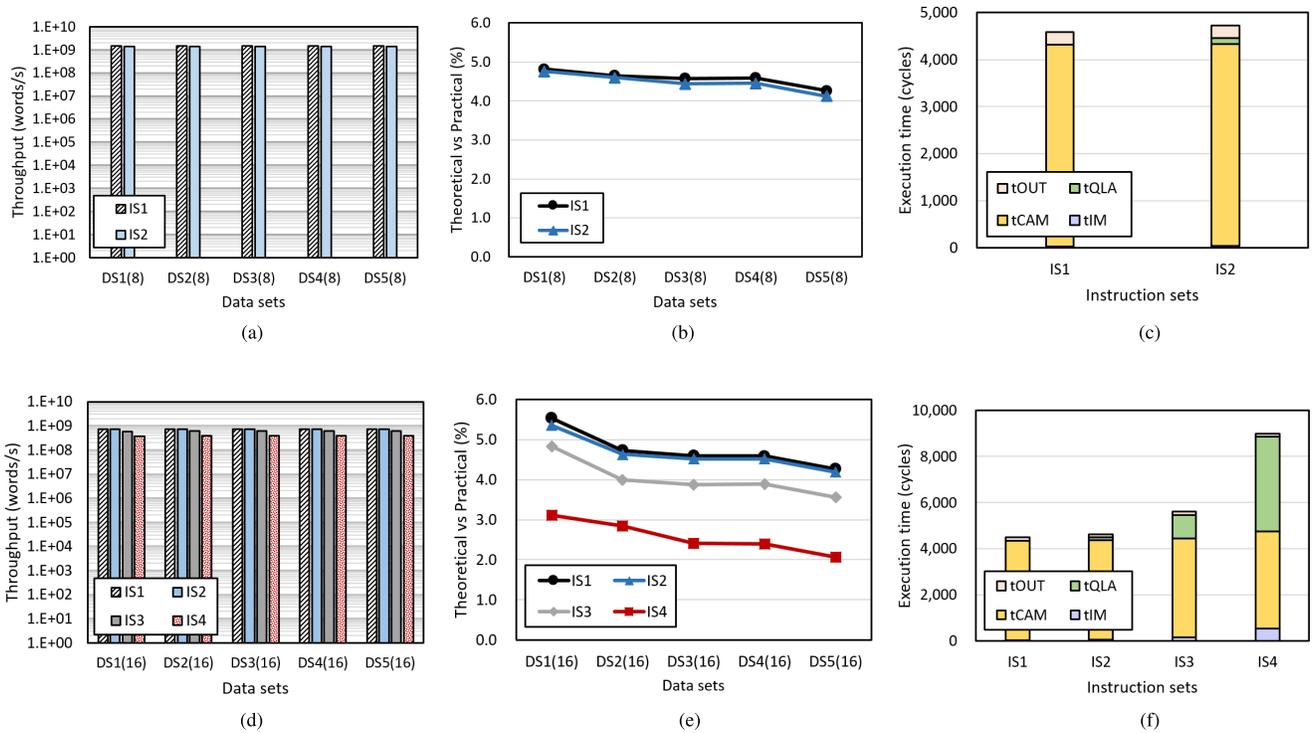

**FIGURE 9.** The results of point-index and full-index experiments. (a) The indexing throughput of BIC64K8. (b) The difference between theoretical and practical throughput of BIC64K8. (c) The time distribution of indexing process of BIC64K8. (d) The indexing throughput of BIC32K16. (e) The difference between theoretical and practical throughput of BIC32K16.

### 3) FULL-INDEX EXPERIMENTS

The full-index experiment is defined as the creation of all available BIs. Specifically, 256 BIs and 65,536 BIs are generated when evaluating the 8-bit and 16-bit data sets, respectively. Taking the 8-bit data set as an example, the indexing process of full-index creation includes: (1) the zeroth batch is copied to BIC64K8, (2) key zero is put into BIC64K8 to obtain the corresponding BI, (3) BI is sent back to the DDR3 memory, (4) the next batch is processed until all of the batches are completed, (5) key one is put into BIC64K8 and the process repeats until 256 keys are indexed. As many as 512 instructions and 131,072 instructions are required to fully index the 8-bit and 16-bit data sets, respectively. Because IM can only store 4,096 instructions at one time, the large instruction sets are divided into 4,096 segments, each of which is processed sequentially. At DS1(8), THR$_{prac}$ reach 90.3 million words/s, which is 3.2% less than THR$_{theo}$. At DS1(16), the THR$_{prac}$ reach 0.37 million words/s, which is 4.3% less than THR$_{theo}$. The difference between THR$_{prac}$ and THR$_{theo}$ is mainly caused by the DDR3 access latency added into t$_{CAM}$ and t$_{OUT}$. Nonetheless, full-index creation is rarely performed in real applications, due to the cost of storage.

### D. ENERGY CONSUMPTION

Fig. 10 shows the comparison of energy consumption (in decimal logarithm) between BIC32K16 and two platforms, which were implemented in CPU [16] and GPU [17], respectively. The energy consumption (J/GB) is calculated as a quotient of power dissipation (W or J/s) and practical indexing throughput (GB/s). The power dissipation parameters of CPU and GPU are extracted from the product specifications provided by the corresponding vendors, while that of BIC32K16 was estimated by the PowerPlay Power Analyzer tool of Quartus 16.0.

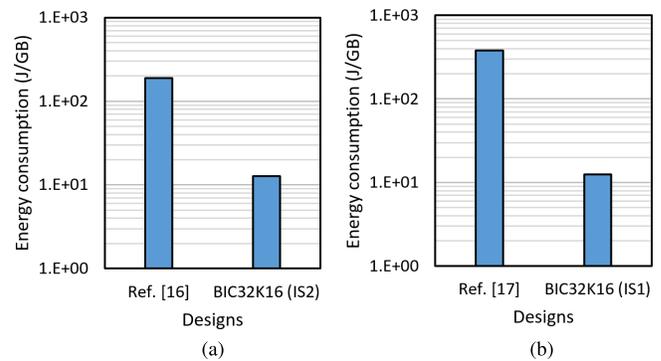

**FIGURE 10.** The comparison of energy consumption. (a) Ref. [16] vs. BIC32K16 (IS2). (b) Ref. [17] vs. BIC32K16 (IS1).

Ref. [16] proposes a CPU-based query system, where all data sets are entirely stored in the DDR3 memories. The data sets included an attribute called *energy*, whose values were always larger than zero. To prepare for the query processing, the authors used Fastbit to generate the BI of (*energy* > 1.2) with a binning precision of two. According to [2] and [25], binning is a special technique that reduces the number of BIs of a high-cardinality attribute by binning some values in this attribute and then producing one bitmap for each bin.

**TABLE 6.** The platform description.

|  | Hardware | Power (W) | Throughput (GB/s) |
|---|---|---|---|
| Ref. [16] | 20,000 cores per 834 Intel CPUs | 95,900 | 510 |
| Ref. [17] | 1,344 cores per 1 NVIDIA GPU | 170 | 0.45 |
| BIC32K16 (IS2) | Intel Arria V FPGA | 18.2 | 1.44 |
| BIC32K16 (IS1) | Intel Arria V FPGA | 18.2 | 1.46 |

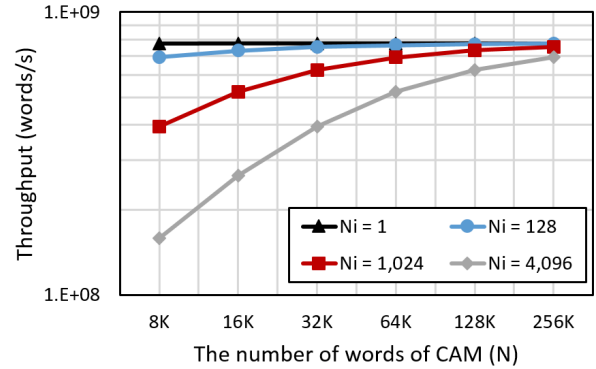

**FIGURE 11.** The variation of indexing throughput.

Moreover, the binning precision of two indicates that the values would be rounded to a two-significant-digit number before being indexed. For example, (*energy* = 1.152) and (*energy* = 1.1527) possess the same BI as (*energy* = 1.15). The indexing throughput reached 510 GB/s in case of 834 Intel CPUs core usage. Since the *energy* is positive, the expression of (*energy* > 1.2) can be rewritten as NOT(*energy* ≤ 1.2). To answer this request, BIC32K16 performs the OR operations of BI of (*energy* = 0.00) to BI of (*energy* = 1.20), and then reverse the result by NOT operation. A total of 123 instructions are needed for this index request. As a result, IS2 is selected for the comparison due to the similarity in the number of instructions usage.

Reference [17] proposed a GPU-based index system, where all data sets were randomly generated and stored in the CPU memory in advance. To prepare for the index processing, data sets were offloaded from CPU to the GPU. As soon as GPU completed the index tasks, the BIs were returned to CPU for further processing. According to the authors, the indexing throughput reached 163 million words/second, which included the transfer time back and forth between CPU and GPU. To make a fair comparison between BIC32K16 and [17], the process time was extracted from the total measured time. The processing throughput of [17] achieved 223 million words/second, or 0.45 GB/s. Since this design produced a point index, IS1 is selected for the comparison.

The summary of hardware utilization and indexing throughput of all designs is shown in Table 6. Specifically, [16] exploited 20,000 cores in 834 Intel CPUs for solving a range index request. Since each CPU consumed 115 W, the total power consumption of 834 CPUs became around 95.9 KW, which led to the energy consumption of 188 J/GB. Reference [17] employed a 1,344-core NVIDIA GPU for indexing a set of random data. Because the GPU required 170-W power, the energy consumption became approximately 377 J/GB. According to the PowerPlay Power Analyzer tool, BIC32K16 consumes 18.2 W in the worst case. Therefore, the energy consumption of BIC32K16 (IS2) and BIC32K16 (IS1) were 12.7 J/GB and 12.4 J/GB, respectively. As seen in Fig. 10, to index 1-GB data, BIC32K16 only requires 6.76% and 3.28% of energy compared to [16] and [17], respectively.

## V. DISCUSSION

Although the proposed BICs outperform CPU- and GPU-based designs in terms of energy efficiency, they have the following restrictions.

• First, the indexing throughput THR$_{prac}$ depends largely on both $t_{CAM}$ and $t_{OUT}$, or specifically the data bus width $w$. In fact, the bottleneck between FPGA chip and DDR3 memory is widely recognized as a major problem in the intensive-bandwidth applications. New FPGA architectures are dealing with this problem. Manish *et al.* [26] suggested an enhanced FPGA-to-DRAM architecture for data center and analytics applications, where memory bandwidth can reach as high as 1,024 GB/s (320 times higher than currently utilized bandwidth of 25.6 Gbps). The performance of BIC, thus, will be improved significantly as new FPGA-DRAM architecture is likely to sharply reduce both $t_{CAM}$ and $t_{OUT}$.

• Second, the ratio of $N$ (the number of R-CAM words) to $N_i$ (the number of instructions) also strongly influences the indexing throughput. Fig. 11 gives the simulation of THR$_{theo}$ based on the formulas in Table 5, when $M = 16$, $N$ ranges from 8,192 (8K) to 262,144 (256K) words and $N_i$ ranges from one to 4,096 instructions. If $N_i$ is much smaller than $N$, e.g., $N_i = 128$ and $N = 64K$, THR$_{theo}$ is almost a straight line, because the domination of $t_{CAM}$ over $t_{QLA}$. On the other hand, if $N_i$ increases, THR$_{theo}$ decreases corresponding to the magnitude of $N$. For example, at $N_i = 4,096$, THR$_{theo}$ drops up to 4.4 times when $N$ varies from 256K to 8K. The maximum THR$_{theo}$ can be broken through, only if the bus width $w$ is extended.

• Third, the size of R-CAM relies heavily on the number of memory blocks of a certain FPGA. Due to the mapping technique, the RAM-CAM ratio reaches 32, or 1-bit CAM costs 32-bit RAM. During the operation, up to 64-Kbit BI is also obtained from R-CAM in parallel, which leads to the huge power consumption in the aspect of the hardware accelerators. This is considered as an inevitable trade-off between performance and power.

• Fourth, the proposed BICs utilized a large number of lookup tables, registers, and embedded memory blocks to maximize the indexing throughput. The current place-and-route compilations were mainly based on the strength of the compiler, especially the optimization setting of aggressive performance mode. However, if the BIC size keeps increasing, the congestion of the routing may severely degrade the performance. One possible solution is to employ the incremental compilation feature provided by Quartus 16.0. This feature allows us to divide a large-sized BIC into sub-partitions, then compile and optimize each partition, and finally combine them into a single project. We can, therefore, achieve the timing closure for a particular partition that does not meet timing requirements, while preserving the compilation results for partitions that have met design requirements. In other words, the incremental compilation can significantly reduce design iteration time and guarantee the whole timing of a large design.

## VI. CONCLUSION

This paper originally exploits the hardware parallelism to construct BIC64K8 and BIC32K16, which can index as many as 65,536 8-bit words and 32,768 16-bit words in parallel, respectively. All processes are fully completed by dedicated hardware to maximize the indexing throughput. When being integrated into a 256-bit system that operates at a 100-MHz frequency, BIC64K8 and a BIC32K16 can index up to 1.43 GB/s and 1.46 GB/s, respectively. Those indexing throughputs were stable, regardless of the size of the data sets. More significantly, the energy consumption of BIC32K16 was 6.76% of CPU-based and 3.28% of GPU-based designs. This achievement confirms the advantages of FPGAs as a prominent solution for energy-efficient compute-intensive emerging applications.

## ACKNOWLEDGMENT

X.-T. Nguyen was with the University of Electro-Communications, Tokyo, Japan.

## REFERENCES

[1] Cisco. *Global Cloud Index: Forecast and Methodology, 2015–2020.* Accessed: Oct. 2017. [Online]. Available: http://www.cisco.com/c/dam/en/us/solutions/collateral/service-provider/global-cloud-index-gci/white-paper-c11-738085.pdf

[2] K. Stockinger and K. Wu, "Bitmap indices for data warehouses," in *Data Warehouses OLAP: Concepts, Architectures Solutions*. Hershey, PA, USA: IGI Global, 2007, pp. 157–178.

[3] R. Karam, R. Puri, S. Ghosh, and S. Bhunia, "Emerging trends in design and applications of memory-based computing and content-addressable memories," *Proc. IEEE*, vol. 103, no. 8, pp. 1311–1330, Aug. 2015.

[4] "Implementing high-speed search applications with Altera CAM," Altera, San Jose, CA, USA, Appl. Note 119, 2001, pp. 1–50.

[5] L. Kyle, "Parameterizable content-addressable memory," Xilinx, San Jose, CA, USA, Appl. Note 1151, 2011, pp. 1–31.

[6] Z. Ullah, K. Ilgon, and S. Baeg, "Hybrid partitioned SRAM-based ternary content addressable memory," *IEEE Trans. Circuits Syst. I, Reg. Papers*, vol. 59, no. 12, pp. 2969–2979, Dec. 2012.

[7] Z. Ullah, M. K. Jaiswal, and R. C. C. Cheung, "E-TCAM: An efficient SRAM-based architecture for TCAM," *Circuits, Syst. Signal Process.*, vol. 33, no. 10, pp. 3123–3144, Oct. 2014.

[8] Z. Ullah, M. K. Jaiswal, and R. C. C. Cheung, "Z-TCAM: An SRAM-based architecture for TCAM," *IEEE Trans. Very Large Scale Integr. (VLSI) Syst.*, vol. 23, no. 2, pp. 402–406, Feb. 2015.

[9] A. Ahmed, K. Park, and S. Baeg, "Resource-efficient SRAM-based ternary content addressable memory," *IEEE Trans. Very Large Scale Integr. Syst.*, vol. 25, no. 4, pp. 1583–1587, Apr. 2017.

[10] A. M. S. Abdelhadi and G. G. F. Lemieux, "Modular SRAM-based binary content-addressable memories," in *Proc. IEEE 23rd Annu. Int. Symp. Field-Programm. Custom Comput. Mach.*, May 2015, pp. 207–214.

[11] H. Wong, H.-F. Liu, F. Olken, D. Rotem, and L. Wong, "Bit transposed files," in *Proc. 11th Int. Conf. Very Large Data Bases*, 1985, pp. 448–457.

[12] P. O'Neil and D. Quass, "Improved query performance with variant indexes," *ACM SIGMOD Rec.*, vol. 26, no. 2, pp. 38–49, 1997.

[13] J. Chou *et al.*, "Parallel index and query for large scale data analysis," in *Proc. Int. Conf. High Perform. Comput., Netw., Storage Anal. (SC)*, 2011, pp. 1–11.

[14] S. Blanas, K. Wu, S. Byna, B. Dong, and A. Shoshani, "Parallel data analysis directly on scientific file formats," in *Proc. ACM SIGMOD Int. Conf. Manage. Data (SIGMOD)*, 2014, pp. 385–396.

[15] G. Chong, H. Li, M. Chen, and M. Zhu, "Accelerate bitmap indexing construction with massive scientific data," in *Proc. Int. Conf. Comput. Sci. Netw. Technol. (ICCSNT)*, 2016, pp. 229–233.

[16] C. Hsuan-Te, J. Chou, V. Vishwanath, and W. Kesheng, "In-memory query system for scientific datasets," in *Proc. IEEE 21st Int. Conf. Parallel Distrib. Syst. (ICPADS)*, Dec. 2015, pp. 362–371.

[17] F. Fusco, M. Vlachos, X. Dimitropoulos, and L. Deri, "Indexing million of packets per second using GPUs," in *Proc. Conf. Internet Meas. Conf. (IMC)*, 2013, pp. 327–332.

[18] X.-T. Nguyen, H.-T. Nguyen, and C.-K. Pham, "An FPGA approach for fast bitmap indexing," *IEICE Electron. Exp.*, vol. 13, no. 4, pp. 1–9, 2016.

[19] X.-T. Nguyen, H.-T. Nguyen, and C.-K. Pham, "A high-throughput and low-power design for bitmap indexing on 65-nm SOTB CMOS process," in *Proc. Int. Conf. IC Design Technol. (ICICDT)*, 2016, pp. 1–4, 2016.

[20] X.-T. Nguyen, H.-T. Nguyen, K. Inoue, O. Shimojo, and C.-K. Pham, "Highly parallel bitmap-index-based regular expression matching for text analytics," in *Proc. IEEE Int. Symp. Circuits Syst. (ISCAS)*, May 2017, pp. 2667–2670.

[21] A. Putnam *et al.*, "A reconfigurable fabric for accelerating large-scale datacenter services," in *Proc. ACM/IEEE 41st Int. Symp. Comput. Architecture (ISCA)*, Oct. 2014, pp. 13–24.

[22] O. Jian, L. Shiding, Q. Wei, W. Yong, Y. Bo, and J. Song, "SDA: Software-defined accelerator for large-scale DNN systems," in *Proc. IEEE Hot Chips Symp. (HCS)*, Aug. 2014, pp. 1–23.

[23] *Arria V SoC Development Kit*. Accessed: Oct. 2017. [Online]. Available: https://www.altera.com/products/boards_and_kits/dev-kits/altera/kit-arria-v-soc.html

[24] *TPC-H Benchmark (Decision Support) Standard Specification Revision 2.17.3*. Accessed: Dec. 2017. [Online]. Available: http://www.tpc.org/tpc_documents_current_versions/current_specifications.asp

[25] *Options for Building Bitmap Indexes*. Accessed: Dec. 2017. [Online]. Available: https://sdm.lbl.gov/fastbit/doc/indexSpec.html

[26] D. Manish, S. Jeffrey, and B. Lance, "Intel stratix 10 MX devices solve the memory bandwidth challenge," Intel, Santa Clara, CA, USA, White Paper, 2016.

[27] X.-T. Nguyen *et al.*, "An efficient FPGA-based database processor for fast database analytics," in *Proc. IEEE Int. Symp. Circuits Syst. (ISCAS)*, May 2016, pp. 1758–1761.

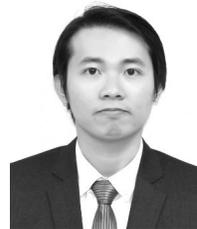

**XUAN-THUAN NGUYEN** (S'15–M'17) received the B.Sc. degree (Hons.) in electronics and telecommunications and the M.Sc. degree in microelectronics and VLSI design from the University of Science, Ho Chi Minh City, Vietnam, in 2010 and 2013, respectively, and the Ph.D. degree from the University of Electro-Communications, Tokyo, Japan, in 2017. He is currently a Post-Doctoral Fellow with the Department of Electrical Computer Engineering, University of Toronto, Toronto, ON, Canada. He was a co-recipient of the Best Student Paper Award Honorable Mention at the 2017 International Symposium on Circuits and Systems Conference. His research interests include energy-efficient hardware approaches to data analytics, artificial intelligence, and embedded systems.

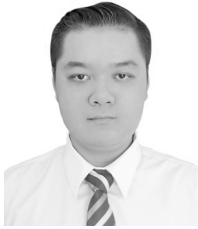

**TRONG-THUC HOANG** (S'17) received the B.Sc. degree in electronics and telecommunications and the M.Sc. degree in microelectronics from the University of Science, Ho Chi Minh City, Vietnam, in 2012 and 2017, respectively. He is currently pursuing the Ph.D. degree with the University of Electro-Communication, Tokyo, Japan. His research field is digital signal processing, image processing, and neural networks.

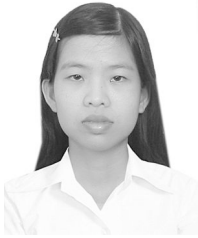

**HONG-THU NGUYEN** (S'14) received the B.Sc. and M.Sc. degrees in electronics and telecommunications from the University of Science, Ho Chi Minh City, Vietnam, in 2011 and 2014, respectively. She is currently pursuing the Ph.D. degree with the University of Electro-Communications, Tokyo, Japan. Her research interests focus on improving communication techniques (MIMO, OFDM, and so on) and designing digital systems using integrated circuits.

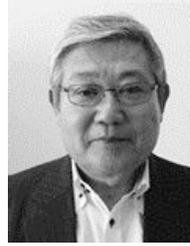

**KATSUMI INOUE** graduated from Tokyo Denki University Tokyo Japan in 1969. He developed Hi-Vision monitors at Ikegami Tsushinki from 1969 to 1973. He established Green Systems in Tokyo for developing micro-computer systems in 1976. He was on the board of UMC Electronics Saitama development division from 1995 to 2007. He established Advanced Original Technologies Co, Ltd. Chiba in 2010. Since 2018, he has been a Ph.D. student at the University of Electro-Communication, Tokyo, Japan. His research interests include the memory-based architecture devices for information detection.

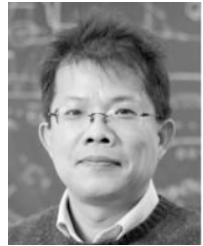

**CONG-KHA PHAM** (M'91) received the B.S., M.S., and Ph.D. degrees in electronics engineering from Sophia University, Tokyo, Japan. He is currently a Professor with the Department of Network Engineering and Informatics, University of Electro-Communications, Tokyo, Japan. His research interests include the design of analog and digital systems using integrated circuits.

• • •